\documentclass[prc,floatfix,nofootinbib,showpacs,showkeys,reprint]{revtex4-1}
\usepackage{graphicx}
\usepackage{amssymb,amsmath,amstext,amsthm,amsfonts}
\usepackage{subfigure}
\usepackage[utf8]{inputenc}
\usepackage{float}
\usepackage{url}

\begin{document}
\today
\title{Muon-neutrino-induced charged-current pion production on nuclei}

\author{U. Mosel}
\email[Contact e-mail: ]{mosel@physik.uni-giessen.de}
\affiliation{Institut f\"ur Theoretische Physik, Universit\"at Giessen, Giessen, Germany}
\author{K. Gallmeister}
\affiliation{Institut f\"ur Theoretische Physik, Johann Wolfgang Goethe-Universit\"at Frankfurt, Frankfurt a.\ M., Germany}

\begin{abstract}
\begin{description}
\item[Background] Long-baseline experiments such as T2K, NOvA or the planned Deep Underground Neutrino Experiment require theoretical descriptions of the complete event in a neutrino-nucleus reaction. Since nuclear targets are used this requires a good understanding of neutrino-nucleus interactions.
\item[Purpose] One of the dominant reaction channels in neutrino-nucleus interactions is pion production. This paper aims for a theoretically consistent treatment of all charged current charged pion production data that are available from the experiments MiniBooNE, the near detector experiment at T2K and MINERvA.
\item[Methods]  Pion production is treated through excitations of nucleon resonances, including background terms, and deep inelastic scattering. The final state interactions of the produced pions are described within the Giessen-Boltzmann-Uehling-Uhlenbeck implementation of quantum-kinetic transport theory.
\item[Results] Results are given for MiniBooNE, the near detector experiment at T2K and for MINERvA. While the theoretical results for MiniBooNE differ from the data both in shape and magnitude, their agreement both with the T2K and the MINERvA data is good for all pion and lepton observables. Predictions for pion spectra are shown for MicroBooNE and NOvA.
\item[Conclusions] Based on the GiBUU model of lepton-nucleus interactions a consistent, good theoretical description of CC charged pion production data from the T2K ND and the MINERvA experiments is possible, without any parameter tunes.
\end{description}
\end{abstract}


\maketitle

\section{Introduction}
The extraction of neutrino-properties from long-baseline experiments, such as T2K, NOvA and the planned Ddeep Underground Neutrino Experiment (DUNE) requires the reconstruction of the incoming neutrino energy from properties of the final state, since unlike in any other nuclear or particle physics experiment the energy of a neutrino beam is not well known. This reconstruction is complicated by the fact that all these experiments use nuclear targets such as H$_2$O (T2K), CH$_2$+Cl (NOvA) or $^{40}$Ar (DUNE). Thus in order to reconstruct the incoming neutrino energy from properties of the final state the neutrino-nucleus interactions have to be well understood.

The initial interactions in such a reaction can be grouped into quasielastic scattering (QE) and many-body 2p2h processes on one hand and inelastic excitations on the other. The latter mainly lead to pion (and other meson) production through nucleon resonances and deep inelastic scattering (DIS). Dedicated experiments to explore the former reactions are MiniBooNE and the T2K near detector (ND) experiment; both these experiments work in a beam of neutrinos with a neutrino energy distribution that peaks at about 0.6 GeV. At about the same energy the experiment MicroBooNE works, that has now started to take data. At the higher energies the experiment MINERvA has contributed significantly to our knowledge of neutrino-nucleus interactions; its incoming flux distribution peaks at about 3.5 GeV while another such experiment, at the NOvA ND, works with a rather narrow flux distribution that peaks at about 2 GeV. The lower-energy experiments MiniBooNE, T2K ND and MicroBooNE all are sensitive only to two of the reaction mechanisms, QE (including 2p2h) and pion production through the $\Delta$ resonance. Pion production there constitutes about 1/3 of the total cross section while at the higher-energy experiments NOvA, MINERvA and, ultimately, DUNE it picks up 2/3 of the total strength \cite{Mosel:2012kt}. It is thus important to have the pion production channels under good, quantitative control. This is essential not only for the actual pion production experiments, but also for so-called 0-pion event samples. The latter are often used for studies of QE and many-body processes; they do not contain any pion in the asymptotic, outgoing state. However, pions (or nucleon resonances) may nevertheless have been produced initially but then reabsorbed. Any analysis of 0-pion events thus necessarily relies also on a good description of pion production (and absorption).

The first extensive neutrino-induced pion production data on nuclear targets were provided by MiniBooNE \cite{AguilarArevalo:2010xt}. Theoretical calculations aiming to describe these data in a state-of-the-art model with a solid foundation in pion-nucleon-$\Delta$ dynamics \cite{Kim:1996bt} all came out to be too low in comparison to the measured cross sections \cite{Leitner:2006ww,Leitner:2008wx,Lalakulich:2011ne,Hernandez:2013jka,Lalakulich:2013tba}. The calculations also predicted a typical shape for the pion kinetic energy distributions, due to pion absorption through the $\Delta$, that was not seen in experiment. More recently, pion production data have also been obtained by the MINERvA experiment \cite{Eberly:2014mra,McGivern:2016bwh}. These data are reasonably well described by the same model that failed for MiniBooNE \cite{Mosel:2014lja}, but there were also marked deviations, in particular at forward angles and low kinetic energies. These remaining discrepancies were tentatively attributed to coherent production in Ref.\ \cite{Mosel:2014lja}.

In a careful comparison of MiniBooNE and MINERvA data the authors of Ref.\ \cite{Sobczyk:2014xza} concluded that the two datasets were not compatible; the authors found noticeable overall normalization problems and a significant discrepancy in the shape of pion kinetic energy spectra. That these two datasets were not compatible was also concluded by the authors of Ref.\ \cite{Alvarez-Ruso:2016ikb}. A recent review of the experimental situation can be found in \cite{Katori:2016yel}.

Most recently, new data on neutrino-induced pion production on water have become available from the T2K ND experiment. This experiment works in a similar energy regime as MiniBooNE. It is, therefore, the aim of this paper to have a theoretical look at these new data and compare them with the MiniBooNE results, with the hope of clarifying the data discrepancies and of establishing a reliable model for pion production on nuclear targets. Using the same theory we will then also calculate pion production spectra for two different sets of data obtained by MINERvA. Finally, we will make predictions for pion spectra in the MicroBooNE and NOvA ND experiments that will soon produce data.

\section{Model}
The model used for these studies is an implementation of transport theory in the form of a generator, dubbed Giessen-Boltzmann-Uehling-Uhlenbeck (GiBUU), that delivers not only inclusive cross sections, but also the full final state event. Its physics content as well as many details of the practical implementation are contained in Ref.\ \cite{Buss:2011mx}. The model approximately factorizes the neutrino-induced pion production on a nucleus into two steps.  First, pion production on a \emph{nucleon} is described by a superposition of a resonance term \cite{Kim:1996bt} and a $t$-channel background term while maintaining the quantum mechanical interference between both amplitudes. The initial interactions with a \emph{nucleus} are then obtained by summing over all interactions with bound and Fermi-moving nucleons.

In a second step then, the final state interactions (fsi) of the produced pions are treated by quantum-kinetic transport theory \cite{Kad-Baym:1962}. This step contains elastic scattering, inelastic resonance excitations, charge transfer from and to the charge channel under consideration and absorption of pions. The complexity of final state interactions necessitates the use of transport theory; it can not be handled within an optical model description such as that used in \cite{Martini:2009uj,Gonzalez-Jimenez:2016glw}. Further details on pion production in GiBUU can be found in Ref.\ \cite{Lalakulich:2010ss}.

While the vector parts of both the resonance and the background amplitudes for pion production on the nucleon are well constrained by electro-pion production data \cite{Drechsel:2007if,Tiator:2009mt} the axial parts are much less known. The calculation of the resonance contribution is well established \cite{Kim:1996bt,Leitner:2009zz}, with some knowledge on the axial $N\Delta$ transition form factors \cite{Hernandez:2007qq,Hernandez:2010bx,Lalakulich:2005cs}. However, there is some uncertainty about possible in-medium changes of the nucleon resonances. Most calculations use a parametrization of the in-medium broadening of the $\Delta$ resonance proposed in Ref.\ \cite{Oset:1987re}. The effects of this broadening on neutrino-induced pion production were investigated in Ref.\ \cite{Lalakulich:2012cj}. There a lowering of the maximal cross section in the MiniBooNE was found when the collisional broadening was turned on. However, it must be noted that the parametrization of the in-medium width of Ref.\ \cite{Oset:1987re} was derived only for rather low momenta of the resonance ($p_\Delta < 0.5$ GeV) and disappears again for higher momenta. Its applicability to the higher-energy experiments is thus doubtful. Therefore, in the present study all calculations were performed without the in-medium broadening of the $\Delta$ resonance.

For the axial background terms there is only little knowledge except from attempts using effective field theory \cite{Hernandez:2007qq,Lalakulich:2010ss}.  A phenomenological prescription then is to calculate the resonance contribution and fit the background term to data for pion production on a nucleon. Such attempts to fit the background amplitude were often hampered by uncertainties about the elementary cross section for neutrino-induced pion production on the nucleon since two available data sets from Argonne National Laboratory (ANL) \cite{Radecky:1981fn} and Brookhaven National Laboratory (BNL) \cite{Kitagaki:1986ct} differed by up to 30 \%. In Ref.\ \cite{Lalakulich:2010ss} an internal inconsistency within the BNL data set was observed. Indeed, later on Ref.\ \cite{Wilkinson:2014yfa,Rodrigues:2016xjj} in a reanalysis of these data, which used the fairly model-independent QE cross section as a benchmark, gave strong indications that the ANL data set was the correct one. In the present calculations we, therefore, use this ANL set to determine the nonresonant contributions to pion production.

The model passes the necessary test of describing the inclusive electron scattering data on nuclear targets such as $^{12}$C. This can be seen by inspecting the resonance region in the comparisons of inclusive cross sections in Ref.\  \cite{Gallmeister:2016dnq}. The actual meson production cross sections in the relevant energy regimes can also reasonably well be described \cite{Krusche:2004uw,Kaskulov:2008ej,ElFassi:2012nr}. At higher energies electron-induced hadronization in nuclei has succesfully been studied in \cite{Gallmeister:2007an}.

The calculations for the various different experiments discussed in this paper are all being done with the very same version of the code (GiBUU 2016)\footnote{The code can be downloaded from \cite{gibuu}.}, without any parameter readjustment, except for target mass and charge. The experiments discussed here contain molecules as targets, such as H$_2$O or CH$_n$. For these we calculate the cross sections for each partner in the molecule and then sum them according to the molecular composition. In order to obtain the cross sections per nucleon we then divide the sum by the total mass number, e.g.\ for a CH target we divide by 13. All cross sections are given per nucleon.

GiBUU results for MiniBooNE and MINERvA had been shown earlier in Refs. \cite{Lalakulich:2012cj,Mosel:2015tja}. The results shown here in this paper are all obtained with an updated version of the theory. As described in \cite{Gallmeister:2016dnq} the new calculations, as far as pion production is concerned, differ in an improved treatment of the nuclear ground state, the use of the free $\Delta$ width, and a different description of the onset of DIS.

\section{Results}

\subsection{MiniBooNE}
\begin{figure}[h]
\centering
\includegraphics[width=0.7\linewidth,angle=-90]{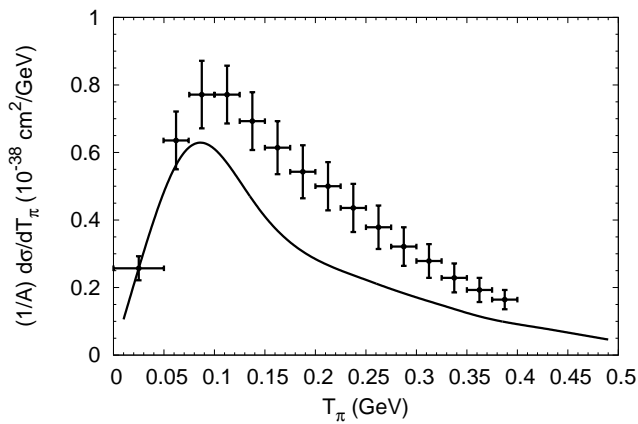}
\caption{Kinetic energy spectrum per nucleon of positively charged pions in the MiniBooNE flux for a CH$_2$ target. Data are from \cite{AguilarArevalo:2010bm}.}
\label{fig:MBpionspectr}
\end{figure}

For completeness we start with only a very short reminder of the results for the MiniBooNE since these have already been extensively discussed \cite{Leitner:2008wx,Lalakulich:2010ss,Lalakulich:2011ne,Lalakulich:2012cj,Hernandez:2013jka,Alvarez-Ruso:2016ikb}. Fig.\ \ref{fig:MBpionspectr} shows the kinetic energy spectrum of positively charged pions in comparison to the MiniBooNE results. It is obvious
that the theory reproduces neither the absolute magnitude of the cross section nor its special shape. The latter is a direct consequence of pion absorption through the $\Delta$ resonance which leads to a decrease of the cross section around a kinetic energy of about 0.18 GeV. At the same time final state interactions acting on the pions lead to an increase below the resonance region thus raising the peak at around 0.08 GeV kinetic energy.
It should be noted that this shape of the pion-kinetic energy distribution has been experimentally observed in photon-induced pion production on nuclei \cite{Krusche:2004uw}. Further results on pion production in MiniBooNE, among them also  energy distributions of the muons as well as angular distributions of the pions and muons, can be found in \cite{Lalakulich:2012cj}.

\subsection{T2K}
The T2K near detector (ND) experiment works in an energy region very close to that of MiniBooNE. In Ref.\ \cite{Gallmeister:2016dnq} we have shown that GiBUU describes the T2K inclusive double-differential cross sections for a $\mu$-neutrino beam very well and thus passes a first, necessary test. Its results on pion production on a water target in T2K \cite{Abe:2016aoo} can thus shed some light on the observed discrepancy of theory and experiment for the MiniBooNE pion data \cite{Lalakulich:2013iaa} even though they may suffer from larger experimental uncertainties \cite{Katori:2016yel}. The T2K pion data were obtained under the kinematical constraints: $p_\mu > 0.2$ GeV, $p_\pi > 0.2$ GeV, cos$(\theta_\mu) > 0.3$, and cos$(\theta_\pi) > 0.3$. These same cuts were also applied in all the calculations.
\begin{figure}[h]
\centering
\includegraphics[width=0.7\linewidth,angle=-90]{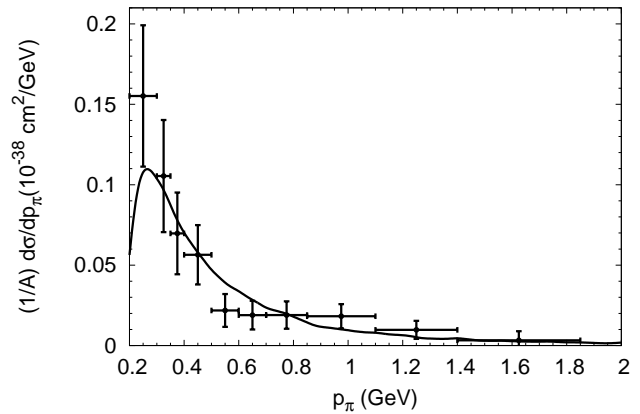}
\caption{Momentum spectrum per nucleon of positively charged pions in the T2K ND280 flux for a H$_2$O target. Data are from \cite{Abe:2016aoo}. For the kinematical cuts used both in the experiment and the calculations see text.}
\label{fig:T2Kpionspectr}
\end{figure}

It is seen in Fig.\ \ref{fig:T2Kpionspectr} that the momentum spectrum of the produced pions is described very well, except for the lowest pion momentum. This result is quite sensitive to the kinematical cuts on both the pion and the muon. A calculation that employs only the cuts on the outgoing $\mu$ gives a perfect description of the spectrum also at the lowest pion momentum. Thus, the pion angular distribution cut $cos(\theta_\pi) > 0.3$ makes the calculated pion momentum distribution to fall below the data at the lowest pion momentum.

\begin{figure}[h]
\centering
\includegraphics[width=0.7\linewidth,angle=-90]{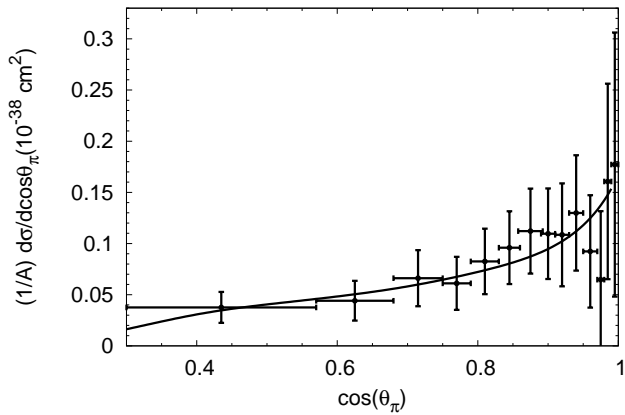}
\caption{Angular distribution of positively charged pions in the T2K ND280 flux for a H$_2$O target. Data are from \cite{Abe:2016aoo}.}
\label{fig:T2Kpionangle}
\end{figure}

The pion angular distribution shown in Fig.\ \ref{fig:T2Kpionangle}, on the other hand, is described very well over the full range of angles; however, at the smallest angles the experimental uncertainties become large.

The discrepancy seen in the spectrum at the lowest pion momentum could be due to three possible effects:
\begin{enumerate}
\item The treatment of pion final state interactions in GiBUU might be responsible for a too strong suppression of slow pions.
\item GiBUU is a semiclassical transport theory that can be applied only to incoherent scattering processes. Coherent neutrino-induced pion production is thus not contained in the theory, but is contained in the data.
\item The observed discrepancy could also be due to details in the $N \Delta$ axial form factor. A slightly different shape with somewhat larger (around 10\%) values around $Q^2 \approx 0.1 - 0.2$ GeV$^2$ could explain the observed cross section.
\end{enumerate}

\begin{figure}[h]
\centering
\includegraphics[width=0.7\linewidth,angle=-90]{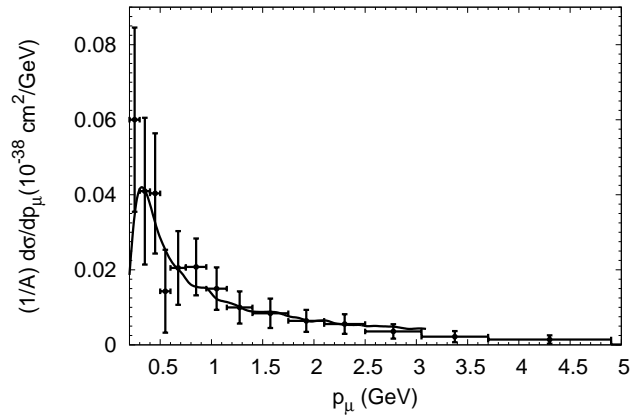}
\caption{Momentum distribution of muons associated with positively charged pions in the T2K ND280 flux for a H$_2$O target. Data are from \cite{Abe:2016aoo}}
\label{fig:T2Kmuspectr}
\end{figure}

\begin{figure}[h]
\centering
\includegraphics[width=0.7\linewidth,angle=-90]{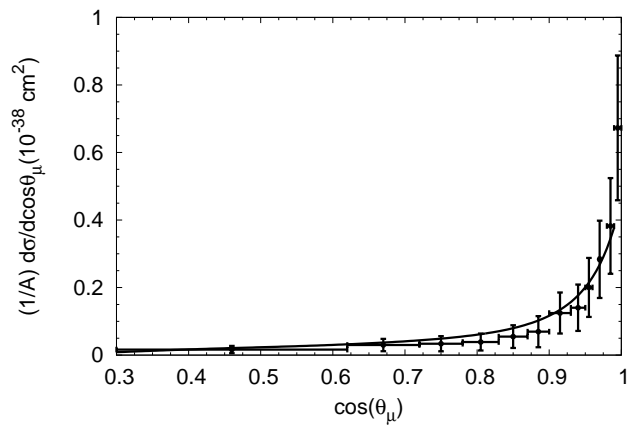}
\caption{Angular distribution of muons connected with positively charged pions in the T2K ND280 flux for a H$_2$O target. Data are from \cite{Abe:2016aoo}}
\label{fig:T2Kmuangle}
\end{figure}

In order to further clarify the reasons for the observed discrepancy at small momenta and angles we now study the outgoing lepton's properties. The distributions of the outgoing $\mu$ lepton (Figs.\ \ref{fig:T2Kmuspectr}, \ref{fig:T2Kmuangle}) show a very similar behavior as for the pion: The agreement of theory with experiment is generally very good, except for the lowest muon momentum and the most forward angle, where the cross sections are slightly underestimated. This indicates that the discrepancy is connected with the smallest $Q^2$. This leaves explanations 2.\ and 3.\ of the list above.

In order to distinguish between these cases a look at experimental data on coherent pion production in the T2K flux is helpful. T2K has indeed measured the coherent $\pi^+$ production, though not on water, but instead on a $^{12}$C target with kinematical cuts that are similar to the ones used above \cite{Abe:2016fic}. The cross section is found to be sizable only for $Q^2 < 0.1$ GeV$^2$ and amounts to about $3 \pm 1 \times 10^{-40}$ cm$^2$. The discrepancy observed in the present calculations is about $0.2 \times 10^{-40}$ cm$^2$ per nucleon, as can be read off Figs.\ \ref{fig:T2Kpionspectr}, \ref{fig:T2Kmuspectr}, and \ref{fig:T2Kmuangle}. This number is compatible with the experimentally determined coherent cross section and thus gives some credence to the assumption that explanation 2.\ above (coherent pion production) is the correct one. However, in \cite{Lalakulich:2013iaa} we have pointed out that the pion production cross section in the T2K experiment is dominated by production through the $\Delta$ resonance; up to a neutrino energy of about 0.8 GeV it proceeds nearly exclusively through this channel. Experimentally distinguishing between the low-$Q^2$ behavior of the $\Delta N$ transition form factor and coherent pion production is thus possible only if it could be verified that the target nucleus remains in its groundstate.

As mentioned earlier, T2K has an incoming energy distribution that is located in a similar energy region as that of MiniBooNE. The fact that the same theory describes the T2K data, but fails to do so for the MiniBooNE data points to an incompatibility of the MiniBooNE to the T2K pion production data.

\subsection{MINERvA}
The MINERvA experiment has obtained the first pion data on a nuclear target (CH) at a higher energy ($\langle E_\nu \rangle$ = 4 GeV) \cite{Eberly:2014mra}. The data were obtained for charged pions, i.e.\ for the sum of $\pi^+$ and $\pi^-$, and they contain a cut on the invariant mass of the initially excited nucleon. The invariant mass can be defined such that it depends only on the incoming gauge boson
$
W_{\rm rec}^2 = M^2 + 2 M \omega - Q^2   ~.
$
Here $M$ is the nucleon mass, $\omega$ the energy transfer and $Q^2$ the squared four-momentum transfer; the nucleon is free and at rest. As pointed out in \cite{Mosel:2015tja} this cut introduces some model-dependence into the data since both $\omega$ and $Q^2$ can not be directly measured, but have to be reconstructed.

The authors of \cite{Eberly:2014mra} complicated this further by first selecting their events by the cut on $W_{\rm rec} < 1.4$ GeV and then correcting this for a cut on a $W < 1.4$ GeV that corresponds to the true invariant mass of a boson-nucleon pair\footnote{The latter step is not trivial since the nucleon is bound and Fermi moving so that $W$ is given by $W^2 = (E_N + \omega)^2 - (\vec{p_N} + \vec{q})^2$. It is not clear to the authors how this could be done within the generator GENIE that does not contain any binding of nucleons.}. Since the actual nucleon is bound and Fermi-moving this $W$ is different from  $W_{\rm rec}$ (see Fig.\ 5 in \cite{Mosel:2015tja}). In addition to the $W$-cut the data also contain a flux cut where the flux is used only between 1.5 and 10 GeV. In the results discussed here both the flux-cut and a $W_{\rm rec}$ cut was taken into account.

\begin{figure}[h]
\centering
\includegraphics[width=0.7\linewidth,angle=-90]{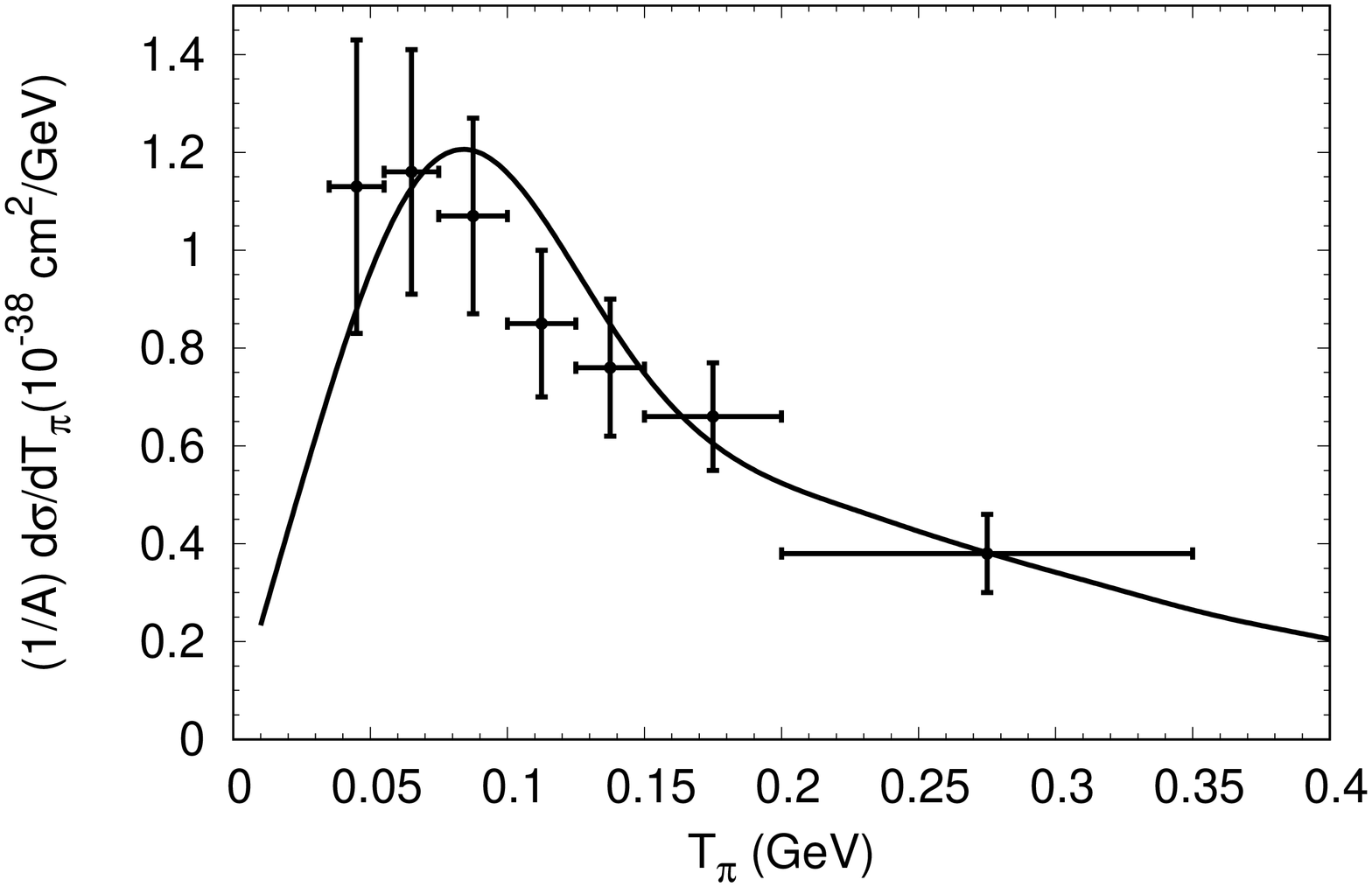}
\caption{Kinetic energy spectrum per nucleon of single charged pions in the MINERvA flux for a CH target with $W < 1.4$ GeV. Data are from \cite{Eberly:2014mra}.}
\label{fig:MINERvApionspectr14}
\end{figure}

\begin{figure}[t]
\centering
\includegraphics[width=0.7\linewidth,angle=-90]{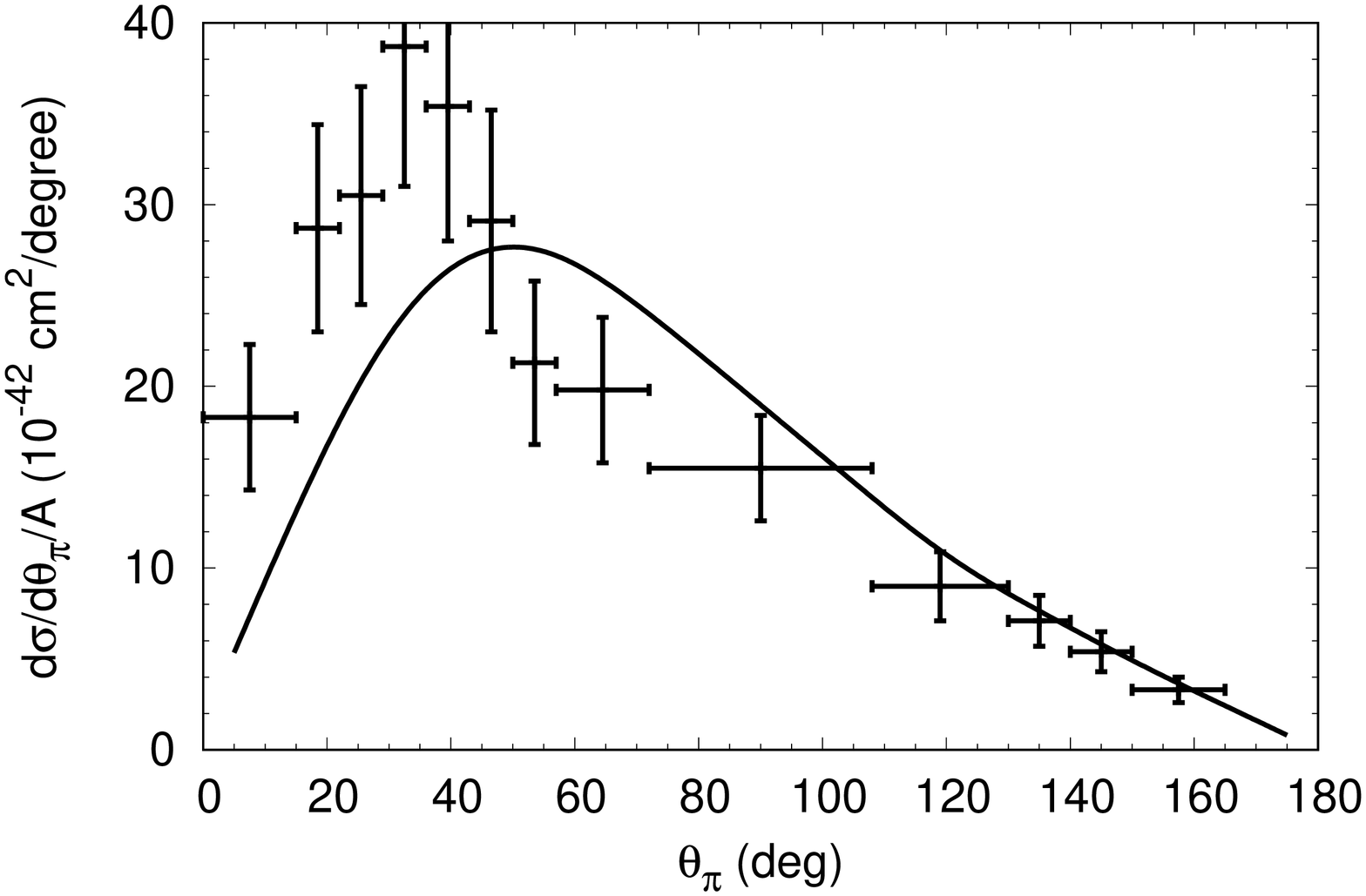}
\caption{Angular distribution of single charged pions in the MINERvA flux for a CH target with $W < 1.4$ GeV. Data are from \cite{Eberly:2014mra}.}
\label{fig:MINERvAKpionangle14}
\end{figure}

Neglecting the difference between the $W_{\rm rec}$- and the $W$-cut we show in Fig.\ \ref{fig:MINERvApionspectr14} a comparison between the calculated  kinetic energy spectrum with the experimental data. The agreement is seen to be quite good, except for the point at lowest energy.
Fig.\ \ref{fig:MINERvAKpionangle14} shows the corresponding angular distribution. Here one observes a clear discrepancy at the angles below about 40$^\circ$.

\begin{figure}[h]
\centering
\includegraphics[width=0.7\linewidth,angle=-90]{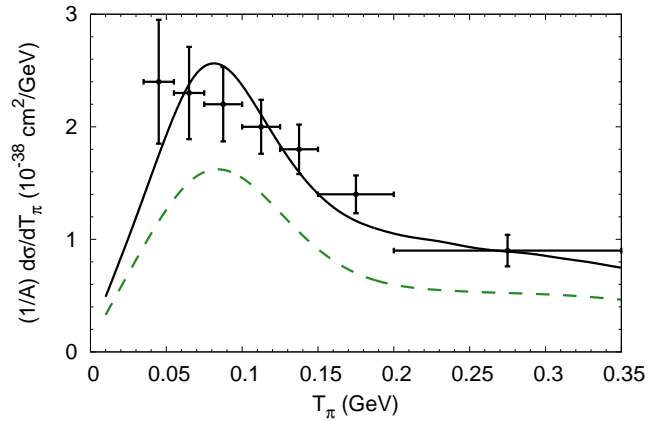}
\caption{Kinetic energy spectrum per nucleon of multiple charged pions in the MINERvA flux for a CH target with $W_{\rm rec} < 1.8$ GeV (solid line). The dashed line gives the 1-pion contribution. Data are from \cite{McGivern:2016bwh}}
\label{fig:MINERvApionspectr}
\end{figure}

More recently the MINERvA collaboration has published a new data set for pion production. In this case the flux was again cut between 1.5 GeV and 10 GeV and a $W_{\rm rec} < 1.8$ GeV cut was imposed. The measured kinetic energy spectrum is compared in Fig.\ \ref{fig:MINERvApionspectr} with the GiBUU result. Generally, the agreement is again very good, with a discrepancy showing up at the lowest energy of $T_\pi = 0.0475$ GeV.
Fig.\ \ref{fig:MINERvApionspectr} also shows by the dashed line the contribution of events with one pion only in the outgoing channel. It is seen that multiple pion events increase that cross section by about 50\%. A closer inspection of the results shows that these multi-pion events are mostly of the type $\pi^+ \pi^0$ and originate to about equal parts from decay of higher-lying nucleon resonances above the $\Delta$ and from DIS.

\begin{figure}[h]
\centering
\includegraphics[width=0.7\linewidth,angle=-90]{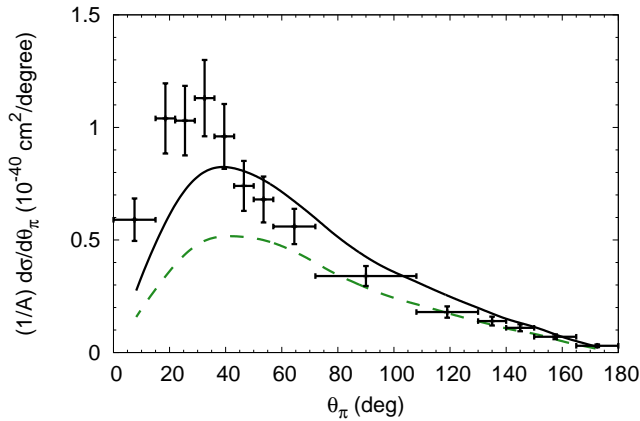}
\caption{Angular distribution  of multiple charged pions in the MINERvA flux for a CH target with $W_{\rm rec} < 1.8$ GeV. The dashed curve gives the angular distribution of 1-pion events. Data are from \cite{McGivern:2016bwh}.}
\label{fig:MINERvApionangle}
\end{figure}

The calculated pion angular distribution is compared with experiment in Fig.\ \ref{fig:MINERvApionangle}. As for the earlier case of the date with the $W < 1.4$ GeV cut there is a noticeable discrepancy at forward angles below about 45$^\circ$.
Again the dashed line gives the contribution of 1-pion events alone.

\begin{figure}[h]
\centering
\includegraphics[width=0.7\linewidth,angle=-90]{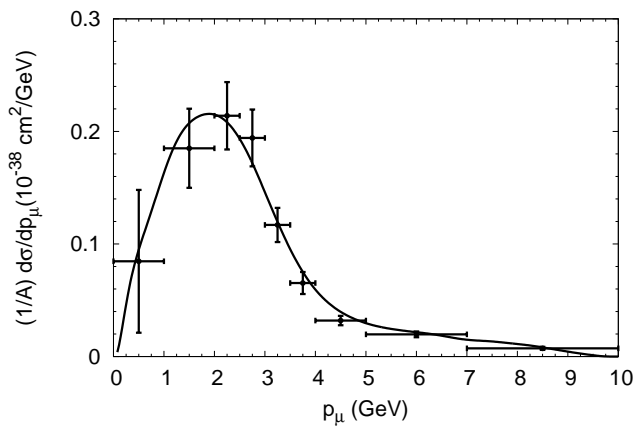}
\caption{Momentum distribution of outgoing muons for multiple charged pion production in the MINERvA flux for a CH target with $W_{\rm rec} < 1.8$ GeV. Data are from \cite{McGivern:2016bwh}.}
\label{fig:MINERvAmuonmomentum}
\end{figure}

\begin{figure}[h]
\centering
\includegraphics[width=0.7\linewidth,angle=-90]{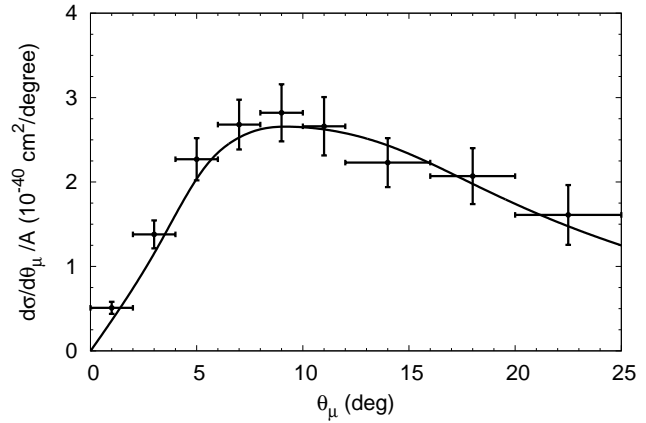}
\caption{Angular distribution of outgoing muons for multiple charged pion production in the MINERvA flux for a CH target with $W_{\rm rec} < 1.8$ GeV. Data are from \cite{McGivern:2016bwh}.}
\label{fig:MINERvAmuonangle}
\end{figure}

Finally, we show in Figs.\ \ref{fig:MINERvAmuonmomentum} and \ref{fig:MINERvAmuonangle} the distributions of the outgoing muons. The agreement is perfect over the whole range of energies. This
shows that the calculations work well also for this semi-inclusive cross section\footnote{When integrating all the differential cross sections shown here for MINERvA one should obtain one and the same total cross section; this can serve as a check for the consistency of the differential ones. We have checked that the calculated integrated differential cross sections indeed all give the same total cross section for pion production.}. Here even at small muon momenta and angles (connected with small $Q^2$) the agreement is very good.

\begin{figure}[h]
\centering
\includegraphics[width=0.7\linewidth,angle=-90]{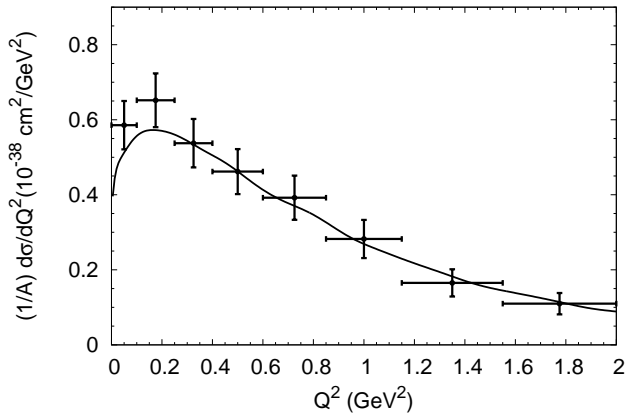}
\caption{$Q^2$ distribution of multiple charged pions in the MINERvA flux for a CH target with $W_{\rm rec} < 1.8$ GeV. Data are from \cite{McGivern:2016bwh}}
\label{fig:MINERvApionQ2}
\end{figure}

In \cite{Mosel:2015tja} it has been argued that the discrepancies seen in the kinetic energy and angular distributions can be due to coherent production which is not contained in the calculated results presented here. Indeed, in \cite{Higuera:2014azj} the coherent cross section in the MINERvA experiment was shown to peak at a pion energy of about 0.4 GeV, i.e.\ at the same energy where the discrepancy appears in Figs.\ \ref{fig:MINERvApionspectr14} and \ref{fig:MINERvApionspectr}. The experimental pion angular distribution for coherent production was found there to peak at about 10$^\circ$, with a slow fall-off towards larger angles which is also compatible with the difference between the experimental and theoretical angular distributions obtained here. The hypothesis that the discrepancies are due to coherent production seems to receive some support from the comparison of the $Q^2$ distributions (Fig.\ \ref{fig:MINERvApionQ2}). It is seen that the agreement with theory is nearly perfect, except for the two lowest points for $Q^2 < 0.2$ GeV$^2$. Coherent production is known to take place in this $Q^2$ region \cite{Higuera:2014azj}. However, it must be remembered that $Q^2$ is not directly measurable so there is some generator dependence in the experimental result.

\subsection{Predictions}
\subsubsection{MicroBooNE}
\begin{figure}[h]
	\centering
	\includegraphics[width=0.7\linewidth,angle=-90]{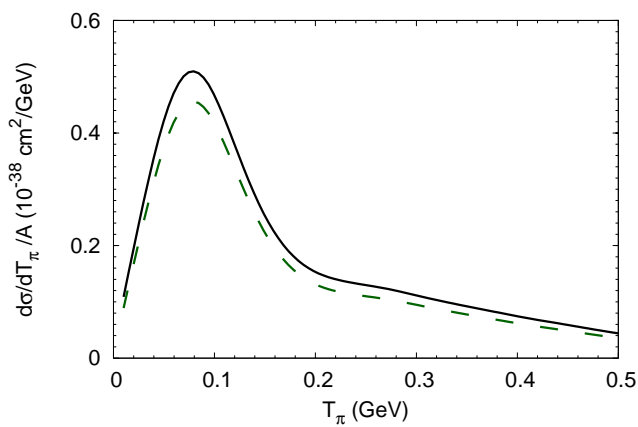}
	\caption{Kinetic energy spectrum of positively charged pions in the BNB flux for a $^{40}$Ar target (MicroBooNE). The dashed
    line gives the spectrum for single $\pi^+$ production, the solid line that for events with at least 1 $\pi^+$.}
	\label{fig:MicroBpionspectr}
\end{figure}

The experiment MicroBooNE works in a flux that is very similar to that of MiniBooNE, but uses a $^{40}$Ar target instead of CH$_2$. In view of the problems discussed above to reproduce the MiniBooNE pion data we show here in Fig.\ \ref{fig:MicroBpionspectr} a prediction for the kinetic energy distribution of CC production of $\pi^+$ in that experiment. Because of the relatively low neutrino energy the pion production is dominated by $1\pi$ events as can be seen by comparing the solid with the dashed line in Fig. \ \ref{fig:MicroBpionspectr}. The shape of the distribution is very similar to that shown in Fig.\ \ref{fig:MBpionspectr} for MiniBooNE.

\subsubsection{NOvA}
The experiment NOvA works in an energy regime in between MicroBooNE and T2K on one side and MINERvA on the other with a relatively narrow beam that peaks at about 2 GeV neutrino energy. Pion production in this experiment could thus supplement our know\-ledge about the transition from $\Delta$ dominated pion production to production through DIS. The calculations were performed for a target with the composition H (10.8\%), C (69.1\%) and Cl (20.1\%) which is close to the actual target composition of H (10.8\%), C (67.0\%), O (2.1\%), Cl (16.9\%), Ti (3.2\%)\footnote{The 2\% contribution of O has been added to C, the 3\% contribution of Ti to Cl. Choosing instead to put all the O and Ti strength into Cl does not change the cross sections per nucleon to any significant degree}.
\begin{figure}[h]
	\centering
	\includegraphics[width=0.7\linewidth,angle=-90]{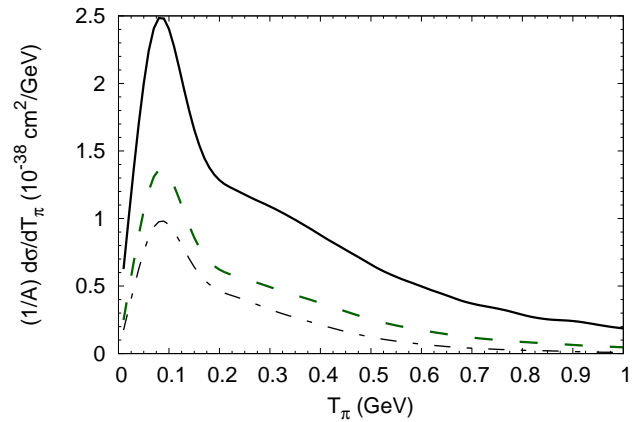}
	\caption{Kinetic energy spectrum per nucleon of positively charged pions in the NOvA flux at the ND. The dashed
    line gives the spectrum for single $\pi^+$ production, the solid line that for events with at least 1 $\pi^+$. The lowest, dashed-dotted curve shows the contribution to single pi production through the $\Delta$ resonance.}
	\label{fig:NOvApionspectr}
\end{figure}

The shape of the kinetic energy distribution is dominated again by a peak at about 0.08 GeV, as in all other experiments shown earlier. This is a feature that is independent of the beam energy.  It reflects the pion absorption through the $\Delta$ resonance that is strongest around $T_\pi = 0.18$ GeV. Pions below that energy cannot easily be absorbed and thus tend to accumulate there. Indeed, pions produced only through the $\Delta$ resonance exhibit exactly the same spectrum as shown by the lowest, dashed-dotted curve in Fig.\ \ref{fig:NOvApionspectr}. The difference between the dashed and the dashed-dotted curve, both for $1\pi$ events, is due to the contribution to this channel from higher-lying resonances as well as to pion absorption events in which one of initially two pions makes it out of the nucleus.

\begin{figure}[h]
	\centering
	\includegraphics[width=0.7\linewidth,angle=-90]{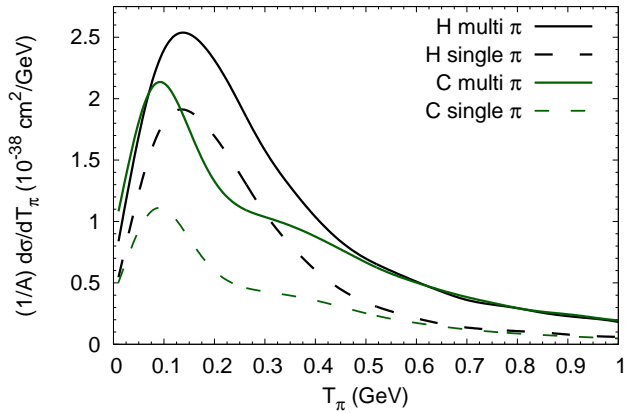}
	\caption{Kinetic energy spectrum per nucleon of positively charged pions in the NOvA flux at the ND. The solid
    lines give the spectrum for events with at least 1 $\pi^+$ plus possibly other mesons, the dashed lines that for events with exactly 1 $\pi^+$.The two upper curves (solid and dashed) describe a H target, the two lower curves a C target.}
	\label{fig:NOvA_HCpionspectr}
\end{figure}

The cross section for multiple $\pi^+$ production is now much larger than that for single pion production, contrary to the situation for MicroBooNE. This just reflects the opening of $2\pi$ decay channels from higher nucleon resonances and the onset of DIS with the higher beam energy, as discussed above for the MINERvA results. In order to illustrate this point further we show in Fig.\ \ref{fig:NOvA_HCpionspectr} the pion spectra for the two dominant target components H and C.
Here the results for H give the pion production cross section before final state interactions, those for C that after fsi.

\section{Conclusions}
In this paper we have compared all the available CC charged-pion data from neutrino-$A$ collisions. The starting point for these investigations was the nagging discrepancy between the MiniBooNE data and theoretical calculations \cite{Lalakulich:2012cj,Hernandez:2013jka} and the seeming incompatibility of the former data with the MINERvA data \cite{Sobczyk:2014xza,Mosel:2015tja,Alvarez-Ruso:2016ikb}.

We have shown that the T2K ND data on a water target agree quite well with the GiBUU calculations, both in absolute magnitude and in shape. The T2K experiment works with a neutrino flux that is centered around the same energy as in the MiniBooNE.  We, therefore, conclude that the latter data are too high and that the shape of the pion kinetic energy distribution determined by MiniBooNE is not correct. This conclusion receives additional support from the observation that all the MINERvA pion data, both the single pion data with a $W< 1.4$ GeV cut and the more recent multiple pion data with a $W < 1.8$ GeV cut, are described very well by the present GiBUU calculations. The remaining discrepancies, mainly at the lowest pion energies and the lowest angles, are compatible with a coherent contribution in the data, which is not contained in the present calculations.

Finally, we stress that all these results were obtained without any tuning of model parameters or changes of model ingredients between experiments. One and the same program version with the same parameter set was used for all the calculations shown in this paper. The good agreement reached with data over different experiments and different neutrino energy regions shows that GiBUU is a reliable tool for the calculation of neutrino-induced pion production.
\begin{acknowledgments}
We gratefully acknowledge the help and support of the whole GiBUU team in developing both the physics and the code used here.

We are also grateful to S.\ Dytman for helpful comments on the MINERvA pion data and to M. Wascko for providing many details about the T2K pion data. J. Paley gave valuable information on the NOvA experiment.

This work has partially been supported by HIC for FAIR.
\end{acknowledgments}

%


\end{document}